\documentstyle[preprint,prb,aps]{revtex}
\begin{document}


\draft

\preprint{Submitted to Physical Review B}

\author{B. D. Rajput and D. A. Browne}

\address{
Department of Physics and Astronomy\\
Louisiana State University\\
Baton Rouge, Louisiana 70803}

\date{\today}

\title{Lattice Dynamics of II-VI materials using adiabatic bond charge
model}

\maketitle

\begin{abstract}
We extend the adiabatic bond charge model, originally developed for
group IV semiconductors and III-V compounds, to study phonons in more
ionic II-VI compounds with a zincblende structure.  Phonon spectra,
density of states and specific heats are calculated for six II-VI
compounds and compared with both experimental data and the results of
other models.  We show that the 6-parameter bond charge model gives a
good description of the lattice dynamics of these materials.  We also
discuss trends in the parameters with respect to the ionicity and
metallicity of these compounds.

\end{abstract}


\narrowtext

\section{Introduction}

The adiabatic bond charge model (BCM) has been quite successful in
explaining the phonon dispersion curves of group IV elemental
semiconductors\cite{Weber} and partially ionic III-V semiconducting
materials\cite{Rustagi} with a zincblende structure.  In recent years
it has been successfully applied to study phonons in semiconducting
superlattices\cite{Yip,Miglio}, optical properties of
Al$_x$Ga$_{1-x}$As\cite{Bernasconi}, open semiconductor
surfaces\cite{Santini} and even sp$^2$-bonded materials like
graphite\cite{Benedek} and fullerenes.\cite{Onida}  Quite recently, a
modified version of the bond charge model has been
applied\cite{Azuhata} to study the second order Raman spectra of AlAs
and AlSb.

In view of the similar dispersion curves of tetrahedrally connected
III-V and II-VI materials, it is surprising that no attempt to extend
BCM to the latter has appeared in the literature.  This may be partly
due to the comments\cite{Kunc&Bilz} indicating that early attempts in
this direction were not successful because it was found\cite{Bilz}
that in the case of II-VI materials the asymmetry of the bond charge
position became too large to find a stable equilibrium position for
the bond charges. These conclusions were based on the studies of the
valence electron charge density using local
pseudopotentials\cite{Walter71}, which suggested a nearly complete
charge transfer from the cation to the anion.  These calculations
actually overestimated the ionicity of these compounds and produced
valence band spectra in strong disagreement with experimental
photoemission results.\cite{Chadi}
Later, more accurate calculations using nonlocal
pseudopotentials\cite{Chelikowski} showed better agreement with the
experiments and yielded charge densities indicating a strong shift of
the bond maximum rather than complete charge transfer.  In fact, the
charge density plots for III-V and II-VI compounds\cite{Chelikowski}
are nearly identical except that the charge maxima in the latter appear
to be slightly shifted toward the anion, indicating that, despite their
greater ionicity, II-VI compounds are dominantly covalent in nature.
In view of these results, it is expected that the BCM should give a
good account of the phonons in II-VI materials provided the bond
charges are placed at physically reasonable places suggested by the
pseudopotential calculations.

Traditionally, the lattice dynamics of these materials has been done
using rigid ion or shell models.\cite{Kunc,Rowe,Talwar,Vagelatos} These
models give good fits to the observed phonon dispersion curves at the
cost of a large number of adjustable parameters (10 or more), some of
which have no physical interpretation.  Recently, {\em ab initio\/}
calculations of phonon spectra have appeared,\cite{Corso} but they are
are not feasible for studying large systems such as alloys like
Cd$_x$Hg$_{1-x}$Te or thick superlattices.  Therefore it is desirable
to have a realistic model with fewer, physically meaningful, parameters
that is easy to extend to more complex systems.  In this paper we show
that the 6-parameter BCM provides a good description of the phonons and
other lattice dynamical quantities such as elastic constants and
specific heat in II-VI materials.

The rest of the paper is organized as follows.  In Sect.~II we provide
a brief overview of the bond charge model.\cite{Weber,Rustagi,Yip}  In
Sect.~III we discuss the results for six II-VI compounds and in
Sect.~IV trends in the parameters are discussed.

\section{Adiabatic Bond Charge Model}

The adiabatic bond charge model (BCM) for homopolar
semiconductors\cite{Weber} and partially ionic III-V
compounds\cite{Rustagi} is the simplest empirical lattice dynamical
model that correctly describes the phonon dispersion curves of covalent
crystals.  In the BCM the valence electron charge density is
represented by massless point particles, the bond charges (BC's), that
follow the ionic motion adiabatically.  The BCM unit cell consists of
two ions and four bond charges that are placed along the bonds between
the ions.  In homopolar covalent crystals the bond charges are placed
midway between the neighboring atoms while in III-V compounds the BC
divides the bond length in the ratio of 5:3.  This is consistent with
nonlocal pseudopotential calculations for the valence electron charge
density\cite{Chelikowski} that indicate that the charge density maximum
in III-V compounds shifts toward the group V element.  This shift is
even stronger in the case of II-VI compounds, reflecting their more
ionic character.  The BCM parameter $p$ which measures the polarity of
the bond is defined in terms of the ratio in which the BC position
divides the bond length.  If $t$ is the bond length and $ r_{1} =
(1+p)t/2 $ and $ r_{2} = (1-p)t/2 $ are the two ion-BC distances then
$p = 0 $ ($ r_{1}/r_{2} = 1 $) for homopolar materials and $ p = 0.25 $
($ r_{1}/r_{2} = 5/3 $) for III-V compounds.  In our extension of the
BCM to II-VI materials we have chosen to use $p = 1/3$ corresponding to
the ratio $r_{1}/r_{2} = 2$ which is based on the
results\cite{Chelikowski} of microscopic calculations.  This choice
will be discussed in more detail in Sect.~\ref{sec:trends}.

The cation and the anion interact with one another and with the bond
charges via central potentials $\phi_{ii}(t)$, $ \phi_{1}(r_1)$, and
$\phi_{2}(r_2)$, respectively.  The bond charges centered on a common
ion interact via a three body Keating potential,\cite{Keating} $
V_{bb}^{\sigma} = B_{\sigma}(\vec
{X}_{i}^{\sigma}{\cdot}\vec{X}_{j}^{\sigma} +
a_{\sigma}^2)^2/8a_{\sigma}^2 $, where $\vec{X}_{i}^\sigma $ is the
distance vector between ion $\sigma$ ($\sigma$ = 1, 2) and BC $i$,
$B_\sigma$ is the force constant and $a_\sigma^2$  is the equilibrium
value of $|\vec{X}_i^\sigma \cdot \vec{X}_j^\sigma|$.  The bond charges
centered on a particular ion also interact directly with one another
through a central potential, $\psi_{\sigma}(r_{bb}^{(\sigma)})$, where
$r_{bb}^{(\sigma)}$ is the distance between the bond charges centered
on the cation ($\sigma = 1$) or the anion ($\sigma = 2$).  Finally, the
ions and the BC's interact via the Coulomb interaction characterized by
a single parameter $Z^2/{\epsilon}$ where $-Ze$ is the charge of a BC,
and $\epsilon$ is the dielectric constant.  Each of the ions is
presumed to have a charge $+2Ze$ so that the net charge in the unit
cell is zero.

To reduce the number of parameters it is assumed\cite{Rustagi} that
$\psi_1' = \psi_2' = 0$,  $\psi_1'' = -\psi_2'' = (B_2 - B_1)/8$ and
$(1+p)\phi_1' + (1-p)\phi_2' = 0$.  If we use these constraints on the
total lattice energy per unit cell
\widetext
\begin{equation}
\Phi = 4[\phi_{ii}(t) + \phi_1(r_1) + \phi_2(r_2)] -
\alpha_{\scriptscriptstyle M} {(2Z)^2 \over \epsilon}{e^2 \over t} +
  6[V_{bb}^1 + V_{bb}^2 + \psi_1(r_{bb}^{(1)}) + \psi_2(r_{bb}^{(2)})]
\end{equation}
along with the equilibrium conditions, ${\partial \Phi/ \partial t} = 0$
and ${\partial \Phi / \partial p}= 0$, we find\cite{error}
\narrowtext
\begin{eqnarray}
\phi_{ii}' & = & -\alpha_{\scriptscriptstyle M}
{Z^2 \over \epsilon}{e^2 \over t^2}\nonumber\\
{\phi_1'\over r_1} & = & 2{d\alpha_{\scriptscriptstyle M} \over dp}
{1-p\over 1+p}{Z^2 \over \epsilon}{ e^2 \over t^3}\nonumber\\
{\phi_2'\over r_2} & = & -2{d\alpha_{\scriptscriptstyle M} \over dp}
{1+p \over 1-p}{Z^2 \over \epsilon}{ e^2 \over t^3}
\end{eqnarray}
The conditions for stable equilibrium,
${\partial^2 \Phi / \partial t^2} > 0$  and
 ${\partial^2\Phi/\partial p^2} > 0$,
further yield
\mediumtext
\begin{equation}
{4 {\phi_{ii}''\over 3} + {(1 + p)}^2 ({\phi_1''\over 3} + {B_2\over 6})
 + {(1 - p)}^2 ({\phi_2''\over 3} + {B_1\over 6}) - {128\over{9\sqrt3}}
\alpha_{\scriptscriptstyle M} {Z^2 \over \epsilon} > 0}
\label{eq:stable1}
\end{equation}
and
\narrowtext
\begin{equation}
{{\phi_1''\over 3} + {\phi_2''\over 3} + {{B_1 + B_2}\over 24} -
{64\over{9\sqrt3}}{d^2\alpha_{\scriptscriptstyle M}\over{dp^2}}
{Z^2 \over \epsilon} > 0}
\label{eq:stable2}
\end{equation}
The Madelung constant $\alpha_{\scriptscriptstyle M}$ of the model is
defined by writing the Coulomb energy per unit cell as
$-\alpha_{\scriptscriptstyle M}(2Ze)^2/\epsilon t$.  For $p=1/3$ the
values of $\alpha_{\scriptscriptstyle M}$, $d\alpha_{\scriptscriptstyle
M}/dp$ and $d^2\alpha_{\scriptscriptstyle M}/dp^2$ are found
numerically\cite{Maradudin} to be  5.0598, 4.0539 and 17.46,
respectively.  In Eq.~(\ref{eq:stable1}) and Eq.~(\ref{eq:stable2}) the
force constants are in units of $e^2/v_a$, where $v_a$ is the unit cell
volume.

With $\phi_{ii}'$, $\phi_1'$, $\phi_2'$, $\psi_1'$, $\psi_2'$,
$\psi_1''$ and $\psi_2''$ given as above, the six free parameters of
the model are $\phi_{ii}''$, $\phi_1''$, $\phi_2''$, $B_1$, $B_2$ and
$Z^2/\epsilon$, which we adjust to fit the neutron scattering data and
the measured elastic constants.  The phonon eigenfrequencies and
eigenvectors are found by diagonalizing the dynamical
matrix\cite{Maradudin} constructed from the BCM equations of motion
\narrowtext
\begin{eqnarray}
{{\bf M}\omega^2{\bf u}} & = & {[{\bf R} + 4{(Ze)^2 \over \epsilon}
 {\bf C}_{\scriptscriptstyle R}]{\bf u} +
[{\bf T} - 2{(Ze)^2 \over \epsilon} {\bf C}_{\scriptscriptstyle T}]
{\bf v}}\nonumber\\
{\bf 0} & = & {[{\bf T}^+ - 2{(Ze)^2 \over \epsilon}
 {\bf C}_{\scriptscriptstyle T}^+]{\bf u} +
[{\bf S} + {(Ze)^2 \over \epsilon}
{\bf C}_{\scriptscriptstyle S}]{\bf v}}
\end{eqnarray}

Here {\bf M} is the mass matrix for the ions and {\bf u} and {\bf v}
are the vectors formed by the displacements of the ions and the BC's,
respectively.  The matrices {\bf R}, {\bf T}, and {\bf S} are the
dynamical matrices for the short range ion-ion, ion-BC and BC-BC
interactions and {\bf C}$_{\scriptscriptstyle R}$, {\bf
C}$_{\scriptscriptstyle T}$ and {\bf C}$_{\scriptscriptstyle S}$ are
the corresponding Coulomb matrices which are evaluated by Ewald's
method.\cite{Maradudin} Explicit forms of {\bf R}, {\bf T}, and {\bf S}
can be found in the appendices of Refs.\onlinecite{Weber} and
\onlinecite{Yip}.

\section{Results and Discussion}

Figure 1 shows the dispersion curves for six II-VI materials along with
the existing neutron scattering data. Figure~2 shows the corresponding
densities of states (DOS).  We present in Table I the BCM parameters
and in Table II the calculated and measured elastic constants.  In all
cases the overall agreement with the experimental data is fairly good
and is of the same quality as for BCM fits for III-V
compounds.\cite{Rustagi} For comparison we have also included the
dispersion curves for GaAs and InSb calculated using the parameters
from Ref.~\onlinecite{Rustagi}.  For CdTe our six parameter fit is as
good as the 14 parameter shell model fit of Rowe {\em et al.\/}\cite
{Rowe} and the 11 parameter rigid ion model fit of Talwar {\em et
al.\/}\cite {Talwar}  All three models show a slight upward bend in the
TO branch in the (100) direction, which is in contrast with the results
of a recent {\em ab initio\/} calculation.\cite{Corso}

The three models give very similar predictions for the phonon density
of states. The principal difference is that the shell model does not
predict a gap in the DOS between the acoustic and optical
contributions, while the BCM has a smaller gap than that seen in the
rigid ion model.  This underscores the fact that the BCM is, in many
ways, an intermediate model between the shell model and the rigid ion
model.  The shell model takes care of the electronic polarizability
explicitly by attaching deformable shells to the ions.  The BCM
partially accounts for the electronic polarizability through the
adiabatic motion of the bond charges, while the rigid ion model ignores
it completely.

For ZnS and ZnTe our fits are comparable with the 10-parameter Valence
Shell Model results of Vagelatos {\em et al.\/}\cite{Vagelatos}  For
ZnTe the BCM predicts a large dispersion in the LO branch near the zone
edge.  However, the maximum deviation from the experimentally measured
frequency at the X point is only about 7 percent.  For both ZnS and
ZnTe the shape of the optical branches in the (110) direction is
different from the results of Ref.~\onlinecite{Vagelatos} but is similar
to that predicted\cite{Corso} by {\em ab initio\/} calculations.  For
ZnSe also the agreement with the neutron data\cite{Hennion} and the
measured elastic constants\cite{Lee} is fairly good.

Mercury compounds, because of their semi-metallic nature, deserve a
separate discussion.  Because of the zero band-gap, the energy for
electronic transitions from the valence band to the conduction band is
comparable to the optical phonon energy.  For HgTe Raman
measurements\cite{Bansal} at 90 K and infrared reflectivity
measurements\cite{Baars} at 77 K yield $\omega_{\scriptscriptstyle
{LO}}\approx138$ cm$^{-1}$ whereas infrared spectra at 8 K
gave\cite{Grynberg} $\omega_{\scriptscriptstyle{LO}}\approx132$
cm$^{-1}$ .  This difference was attributed to the large number of
carriers at higher temperatures.  The same reason is invoked to explain
the degenerate values of $\omega_{\scriptscriptstyle{LO}}$ and
$\omega_{\scriptscriptstyle{TO}}$ found in neutron scattering
experiments.\cite{Kepa}  Because of this controversy we did not use
optical phonon frequencies near the zone center in our fit for HgTe.
It is seen that the agreement with the acoustic and transverse optical
phonons is good.  However, the fit for the LO branch is not of the same
quality, although the deviation from the experimental points is only a
few percent.  In the (111) direction the BCM predicts that the LO
branch will dip downward instead of the upward trend observed
experimentally.\cite{Kepa} For HgSe the agreement with the available
neutron data on acoustic phonons\cite{Kepa} and optical
measurements\cite{Witowski} and measured elastic constants\cite{Ford}
is very good.  Because of the lack of neutron data on optical phonons
we cannot comment on the accuracy of the optical branches.  However, it
should be mentioned that the BCM and the 11 parameter rigid ion
model\cite{Kepa} give similar behavior for the optical branches.

To further check the parameters we have calculated the specific
heats\cite{Maradudin} for all the six materials using the parameters
given in Table I.  The results are shown in Fig.~3 as plots of log $C$
{\em vs.\/} log $T$ along with the experimental data.  In every case
good agreement is obtained with the experiments, giving further support
for the parameters used and the calculated density of states.

We have thus demonstrated that the 6-parameter BCM provides a good
description of the phonons and other lattice dynamical quantities such
as elastic constants and specific heat for II-VI compounds with
zincblende coordination.  The overall agreement with the neutron data
is very good.  However, some discrepancies remain, particularly in the
LO branch near the zone edge, where the BCM predicts a large dispersion
in almost every material including the III-V compounds.  This is most
obvious in HgTe and is indicative of the failure of the BCM to account
for the polarizability of the ion core.  Because of the associated
macroscopic field, the LO phonons are more affected than the other
branches.  The calculated\cite{Walter} static dielectric function
$\epsilon(q)$ for III-V and II-VI compounds is known to have
considerable structure, so including a charge form factor should
remedy this discrepancy.\cite{Bilz}

\section{Trends in Parameters}\label{sec:trends}

Some trends in the parameters presented in Table I are immediately
obvious.  One notices that as one goes from group IV elements to III-V
compounds to II-VI compounds the parameters involving BC's change
uniformly.  For group IV elements the bond charge is situated midway
along the bond and the ion-BC and BC-ion-BC force constants are equal
for the two ions.  However, in III-V compounds the BC shifts toward the
anion which results in higher values for $\phi_2''/3$ and $B_2$ than
$\phi_1''/3$ and $B_1$ respectively.  This trend continues as we move
to II-VI compounds in which the BC is even closer to the anion.  This
pattern in the values of these parameters can be traced to the
equations linking $\phi_1'$ and $\phi_2'$ to
$\alpha_{\scriptscriptstyle M}$ and $d\alpha_{\scriptscriptstyle
M}/dp$.  It should be noted that the values for
$\alpha_{\scriptscriptstyle M}$ and $d\alpha_{\scriptscriptstyle M}/dp$
for II-VI compounds are higher than those for III-V compounds.  Apart
from these obvious features, there are no other discernible trends in
the parameters with respect to the ionicity or the bond length.
However, it is seen that the ion-bond parameters $\phi_1''/3$ and
$\phi_2''/3$ are considerably lower for mercury compounds than for
other materials.  This is reasonable in view of the semi-metallic
nature of these materials and the fact that these parameters represent
off-diagonal contributions to the dielectric function.

The effect of ionicity on the phonon dispersion curves can be
investigated by studying the isoelectronic sequence of materials in
which the bond lengths and the average mass in the unit cell are almost
same.  Two such sequences are Ge-GaAs-ZnSe and $\alpha$Sn-InSb-CdTe.
Increased ionicity results in a general lowering of all frequencies and
elastic constants and a lifting of the LO-TO degeneracy at the $\Gamma$
point and the LO-LA degeneracy at the X point.  A glance at the BCM
parameters for these materials shows that the only pattern is a
decrease in the magnitude of the bond charge $Z$ and, in the case of
Ge-GaAs-ZnSe, a decrease in $\phi_{ii}''/3$ with increased ionicity.
For the other sequence,  $\phi_{ii}''/3$ is almost same for $\alpha$-Sn
and InSb but it is smaller for CdTe.  A cross comparison of the
corresponding materials in the two sequences shows that moving down
the periodic table, with its concomitant increase in metallicity,
yields an increase in the ion-ion interaction $\phi_{ii}''/3$ while the
parameters involving the bond charges decrease.

We should also comment on the choice of the equilibrium positions of
the bond charges.  In principle, $p$ should be treated as the seventh
adjustable parameter of the model.  However, we decided to use the
physically reasonable value of 1/3 for $p$.  This corresponds to
dividing the bond length in a ratio 2:1 and is consistent with the
pseudo-potential calculations of the valence electron charge
density.\cite{Chelikowski}  However, we were also able to find values
for the six parameters which still satisfied the stability conditions
(\ref{eq:stable1}) and (\ref{eq:stable2}) and which gave satisfactory
fits for $p$ as high as 0.55.  The parameters that varied most with $p$
were the ion-BC force constants; increasing $p$ led to a larger
$\phi_2''/3$ and a smaller $\phi_1''/3$.  The remaining parameters also
changed, but by much less.  In every case, the acoustic phonon curves
showed very good agreement with the neutron scattering data.  However,
the agreement with the optical phonons slightly worsened as $p$
increased.  These results highlight the arbitrariness involved in
defining the equilibrium position for the bond charges.  Our choice of
$p = 1/3$, which coincides roughly with the position predicted by the
pseudopotential calculations\cite{Chelikowski}, still gave the best
overall agreement with the experimental results.

\section{Summary and Conclusion}

We have applied the adiabatic bond charge model to study phonons in six
II-VI compounds with a zincblende structure.  The theoretical
predictions of the 6-parameter BCM are in good agreement with the
available neutron data and the experimentally measured elastic
constants and specific heats.  Some minor discrepancies in the LO
branch near the zone edge are believed to be due to the incomplete
description of the electronic polarizability of the ions.  These
deviations are larger, though still only a few percent, in the case of
HgTe as expected from its semi-metallic nature and consequently
stronger screening effects.  In conclusion, we have found that the
6-parameter adiabatic bond charge model provides a satisfactory
description of the lattice dynamics of tetrahedrally connected II-VI
compounds.  The agreement with the experimental data is of the same
quality as for III-V compounds.

We have also discussed some broad trends seen in the parameters.  We
find that the parameters for the bond charge-cation short range
interactions decrease with a corresponding increase in the parameters
for the bond charge-anion interactions as one goes from group IV
elements to III-V to II-VI compounds.

We thank T. Golding for useful conversations.  This work was supported
by the National Science Foundation under grant No.~NSF-DMR-9408634.

\begin{figure}

\caption{
Calculated phonon dispersion curves for CdTe, ZnS, ZnTe, ZnSe, HgTe,
HgSe, GaAs, and InSb.  The BCM Parameters for GaAs and InSb were taken
from Ref.~\protect\onlinecite{Yip}.  Empty circles indicate neutron
scattering data taken from Refs.~\protect\onlinecite{Rowe} (CdTe),
\protect\onlinecite{Vagelatos} (ZnS and ZnTe),
\protect\onlinecite{Hennion} (ZnSe),
 \protect\onlinecite{Kepa} (HgSe and
HgTe), \protect\onlinecite{Waugh} (GaAs), and
\protect\onlinecite{Price} (InSb).}

\end{figure}

\begin{figure}

\caption{
Phonon density of states for II-VI compounds calculated using the
root sampling method.  The fine structure on the curves is an
artifact of the numerical method.}

\end{figure}

\begin{figure}

\caption{
Log $C$ $vs$ log $T$ plots for the calculated and measured specific
heats of several II-VI materials.  The experimental data is taken from
Refs.~\protect\onlinecite{Irwin,Birch,Collins,TalwarV}.}

\end{figure}

\mediumtext

\begin{table}

\caption{BCM parameters for group IV elements, III-V and II-VI
compounds.  Force constants are in units of $e^2/v_a$, where $v_a$
is the unit-cell volume.
\label{table1}
}

\begin{tabular}{cccccccc}
&$\phi_{ii}''/3$  &$\phi_1''/3$ &$\phi_2''/3$
&$B_1$  &$B_2$  &$Z^2/\epsilon$  &$Z$\tablenotemark[3]\\
\tableline
Si\tablenotemark[1]& 6.21 & 6.47 & 6.47 & 8.60 & 8.60 &
0.1800 & 1.47 \\
Ge\tablenotemark[1]& 6.61 & 5.71 & 5.71 & 8.40 & 8.40 &
0.1620 & 1.61 \\
$\alpha$-Sn\tablenotemark[1]& 7.43 & 5.59 & 5.59 & 7.80 &
7.800  & 0.163 & 1.98 \\ \\

AlAs\tablenotemark[2]& 5.80 & 2.27 & 15.48 & 5.79 & 8.54 &
0.1800 & 1.21\\
GaP\tablenotemark[2]& 6.04 & 2.4 & 17.91 & 5.20 & 10.0 &
0.2030 & 1.36\\
GaAs\tablenotemark[2]& 6.16 & 2.36 & 16.05 & 5.36 & 8.24 &
0.1870 & 1.43 \\
GaSb\tablenotemark[2]& 6.77 & 2.37 & 13.10 & 6.28 & 7.08 &
0.1600 & 1.52\\
InP\tablenotemark[2]& 7.16 & 2.95 & 21.62 & 3.43 & 8.37 &
0.2490 & 1.55\\
InAs\tablenotemark[2]& 7.31 & 2.64 & 17.86 & 3.99 & 7.30 &
0.2100 & 1.60\\
InSb\tablenotemark[2]& 7.47 & 2.33 & 14.09 & 4.56 & 6.24 &
0.1720 & 1.64 \\ \\

ZnS& 5.74 & 0.79 & 29.90 & 0.83 & 15.40 & 0.2130 & 1.05 \\
ZnSe& 5.01 & 1.19 & 22.82 & 1.21 & 15.65 & 0.1790 & 1.03 \\
ZnTe& 5.51 & 1.06 & 22.93 & 1.07 & 17.00 & 0.1800 & 1.05 \\
CdTe& 6.85 & 0.77 & 23.34 & 0.39 & 15.44 & 0.1830 & 1.15 \\
HgSe& 5.32 & 0.15 & 14.01 & 0.35 & 17.50 & 0.1095 & 0.91\\
HgTe& 6.46 & 0.081 & 13.46 & 1.08 & 15.60 & 0.1062 & 1.03\\
\end{tabular}
\tablenotetext[1]{Parameters from Ref.~\onlinecite{Weber}.}
\tablenotetext[2]{Parameters from Ref.~\onlinecite{Yip}.}
\tablenotetext[3]{$\epsilon_\infty$ for HgSe from
Ref.~\onlinecite{Einfeldt},
for HgTe from Ref.~\onlinecite{Grynberg} and for the rest
of the materials from Ref.~\onlinecite{Kunc2}.}

\end{table}

\narrowtext

\begin{table}
\caption{Theoretical and measured values (in parantheses)
for the elastic constants in units of $10^{11}$ dynes/cm$^2$.
\label{table2}}
\begin{tabular}{cccc}
&c$_{11}$ &c$_{12}$ &c$_{44}$\\
\tableline
ZnS\tablenotemark[1]& 10.907 (10.46) & 6.498 (6.53) & 4.678 (4.61) \\
ZnSe\tablenotemark[2]& 8.996 (8.59) & 5.064 (5.06) & 4.056 (4.06) \\
ZnTe\tablenotemark[1]& 7.138 (7.13) & 4.233 (4.07) & 3.122 (3.12) \\
CdTe\tablenotemark[1]& 5.675 (5.35) & 4.073 (3.68) & 2.047 (1.994) \\
HgSe\tablenotemark[3]& 6.218 (6.22) & 4.647 (4.64) & 2.262 (2.27) \\
HgTe\tablenotemark[4]& 5.631 (5.631) & 3.785 (3.66) & 2.123 (2.123) \\
\end{tabular}
\tablenotetext[1]{Measured values from Ref.~\onlinecite{Berlincourt}.}
\tablenotetext[2]{Measured values from Ref.~\onlinecite{Lee}.}
\tablenotetext[3]{Measured values from Ref.~\onlinecite{Ford}.}
\tablenotetext[4]{Measured values from Ref.~\onlinecite{Cottam}.}

\end{table}


\begin{references}
\bibitem{error}
There are a couple of misprints in Ref.~\onlinecite{Yip}.
\end{references}
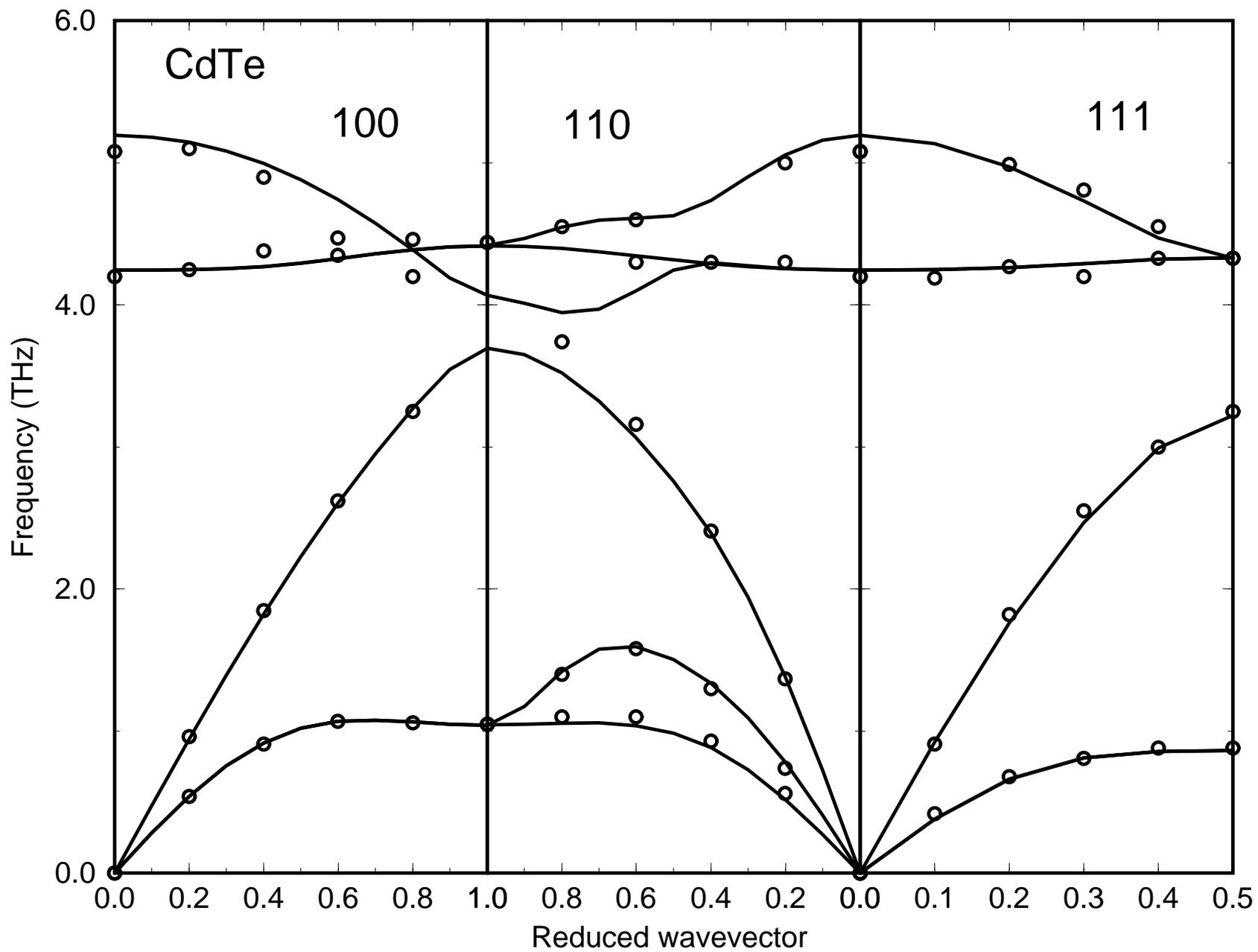
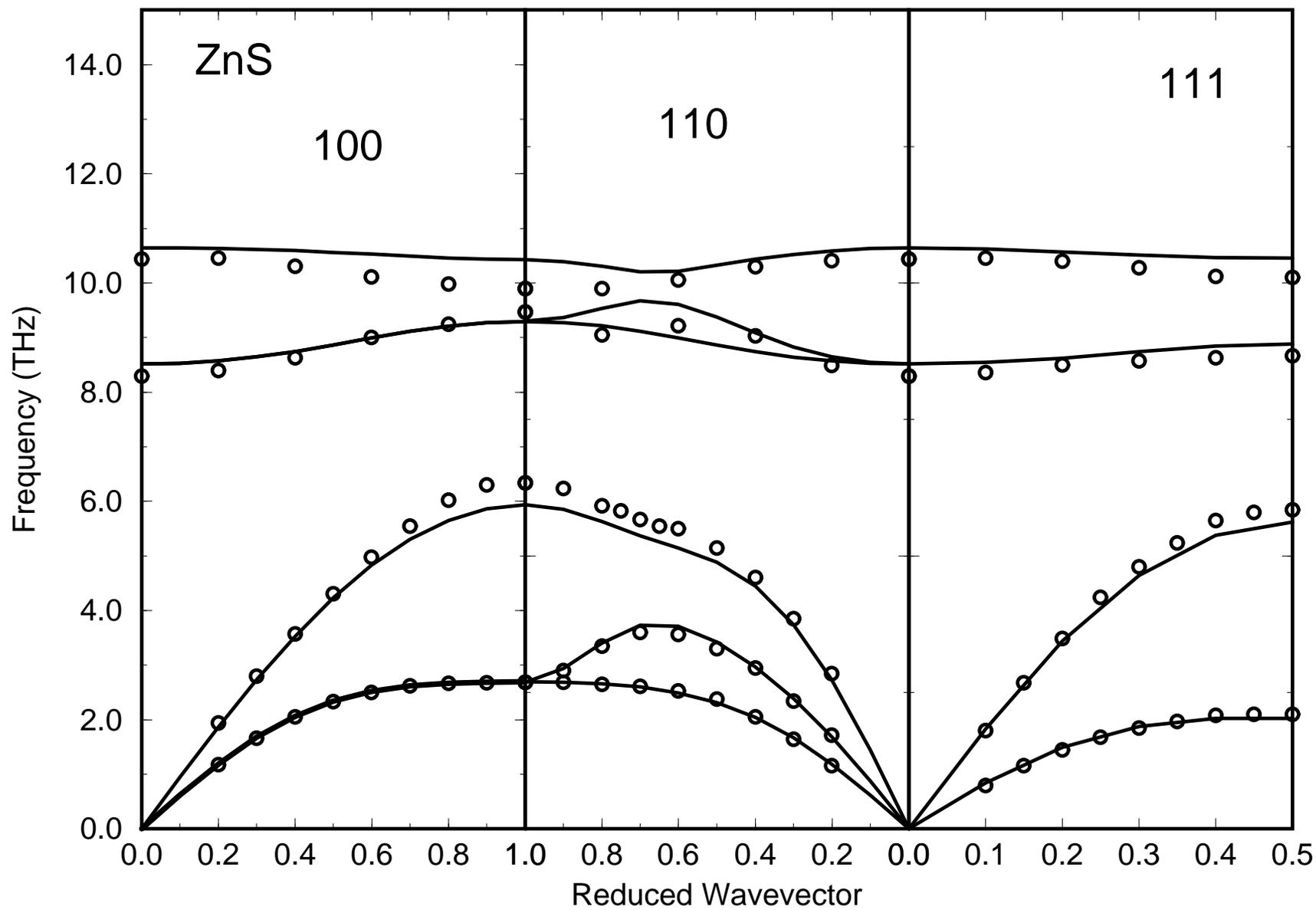
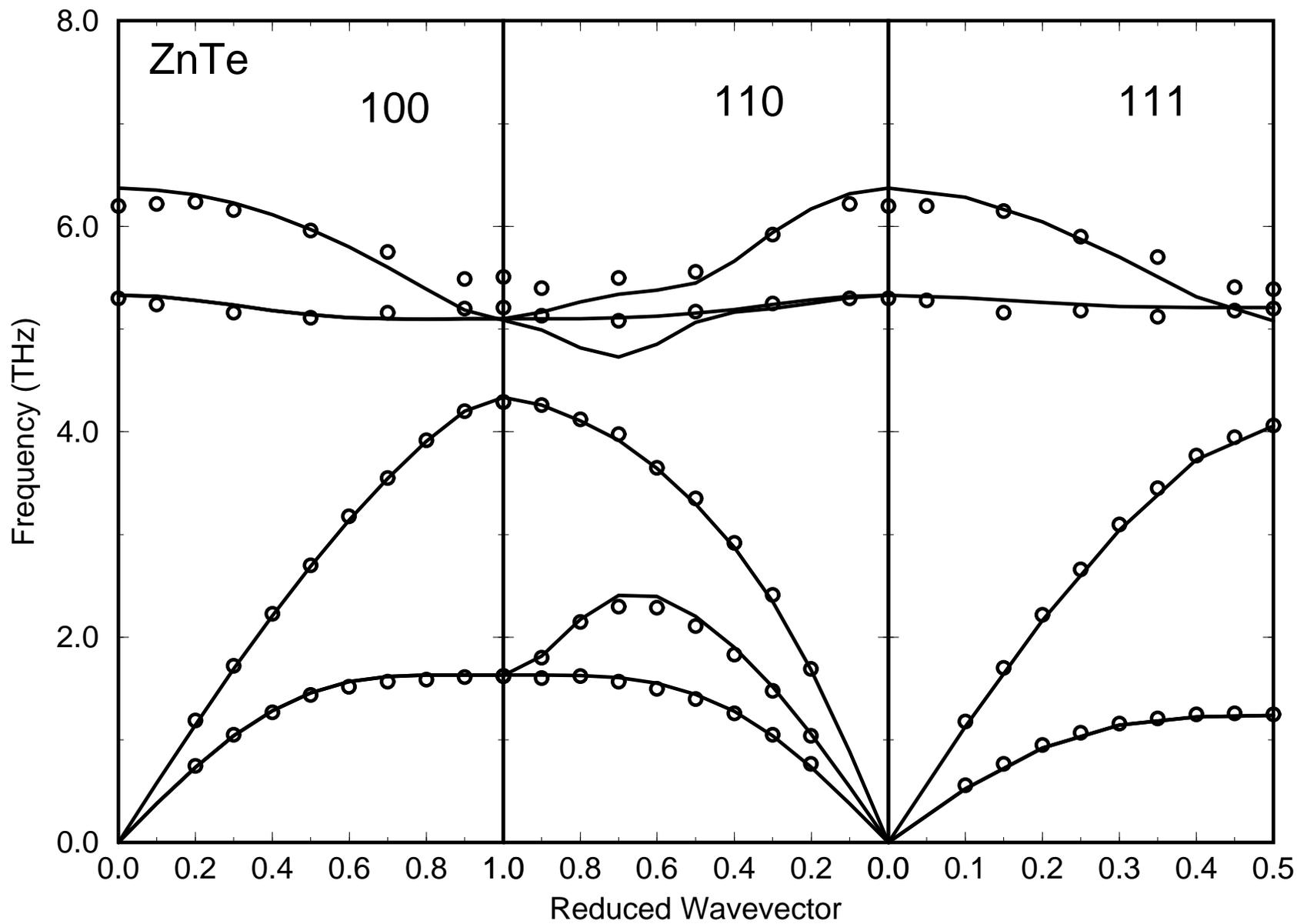
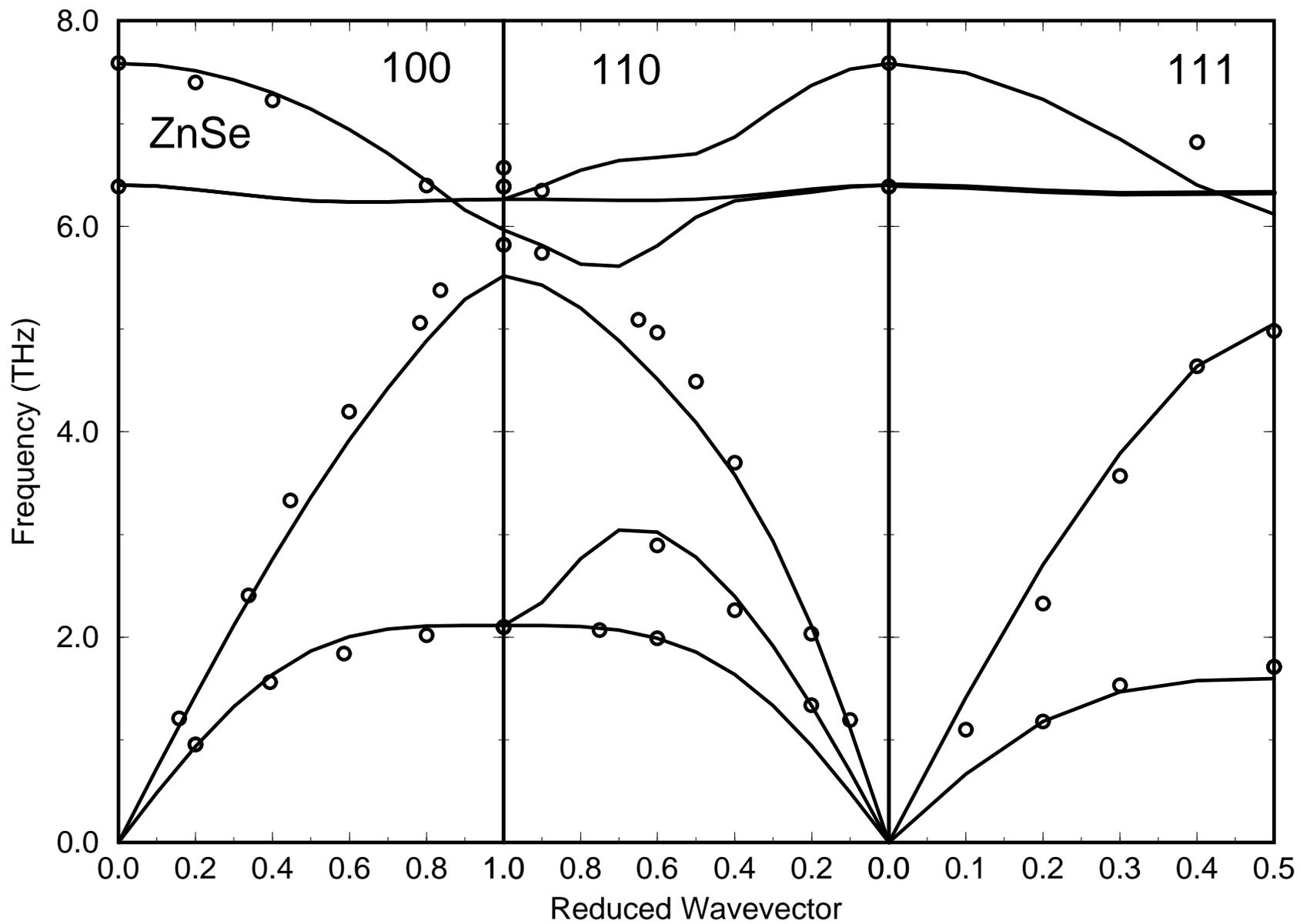
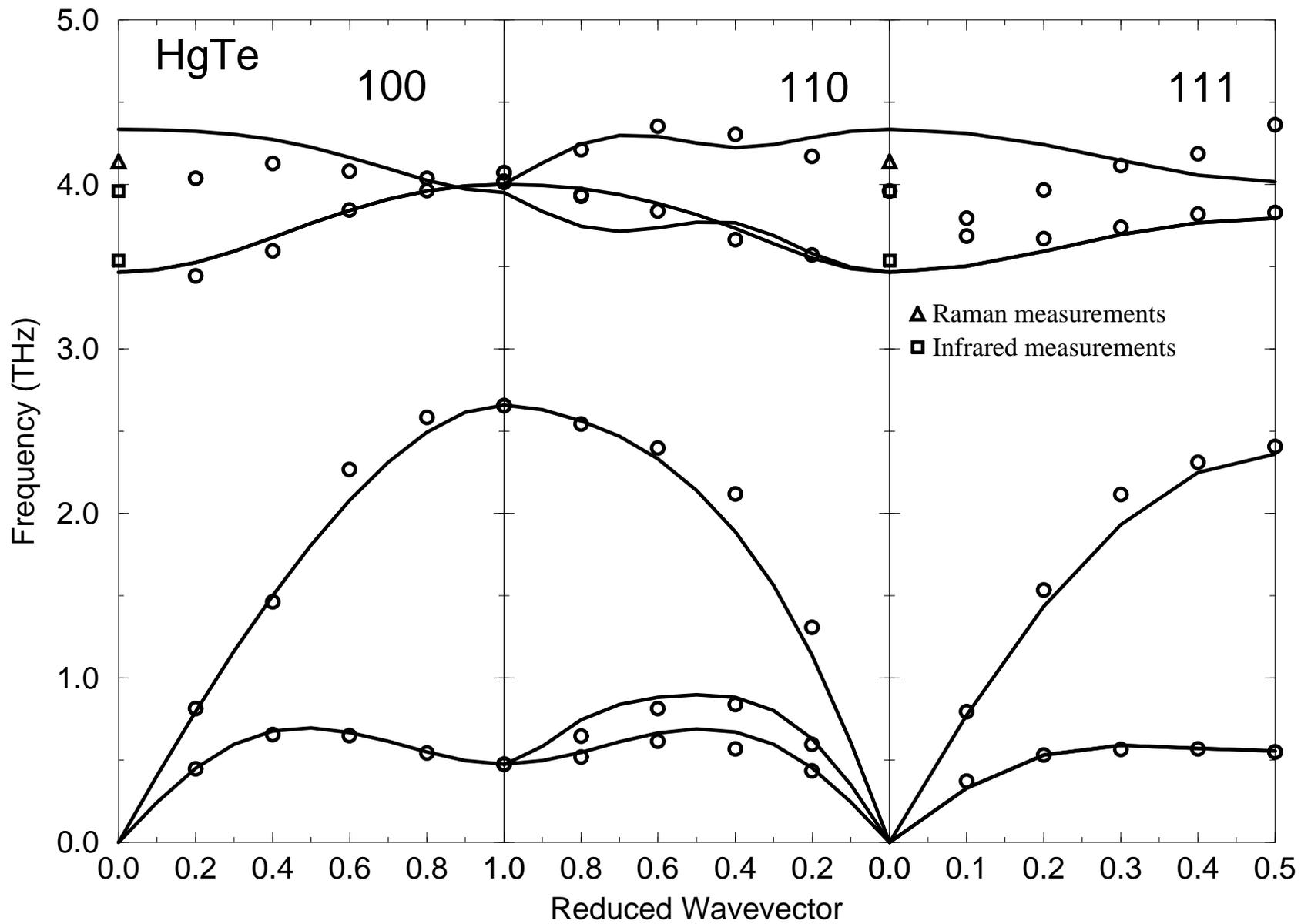
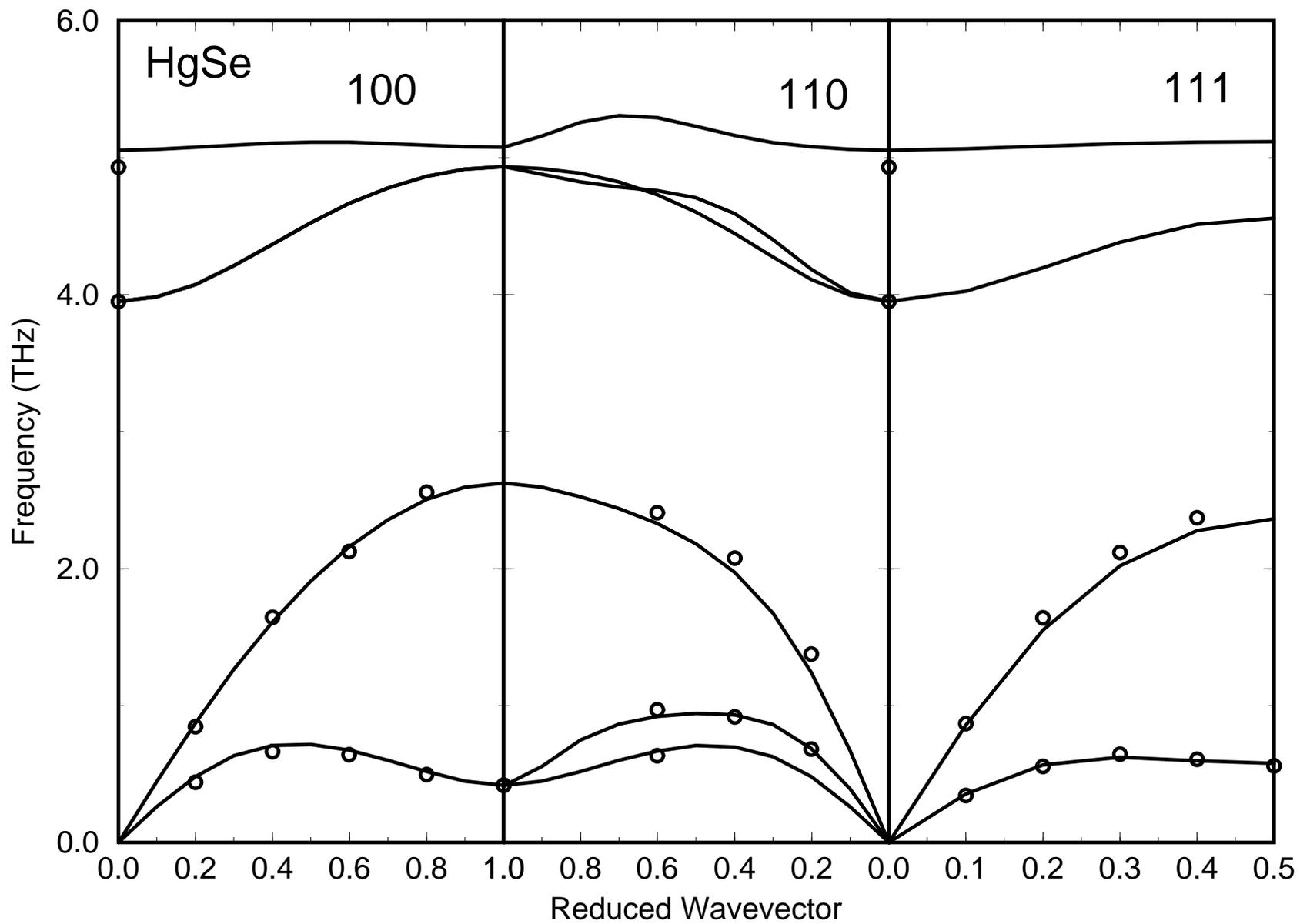
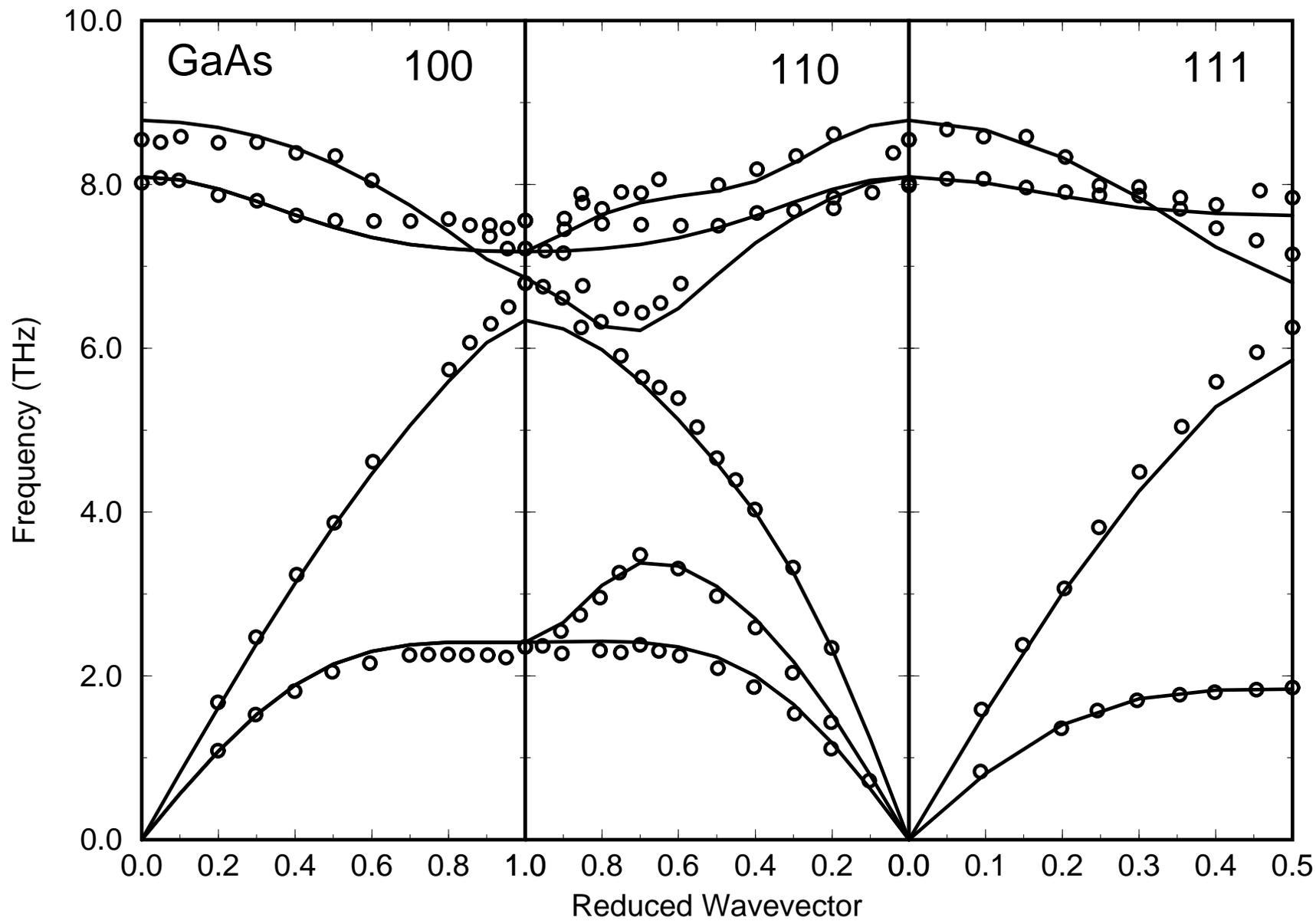
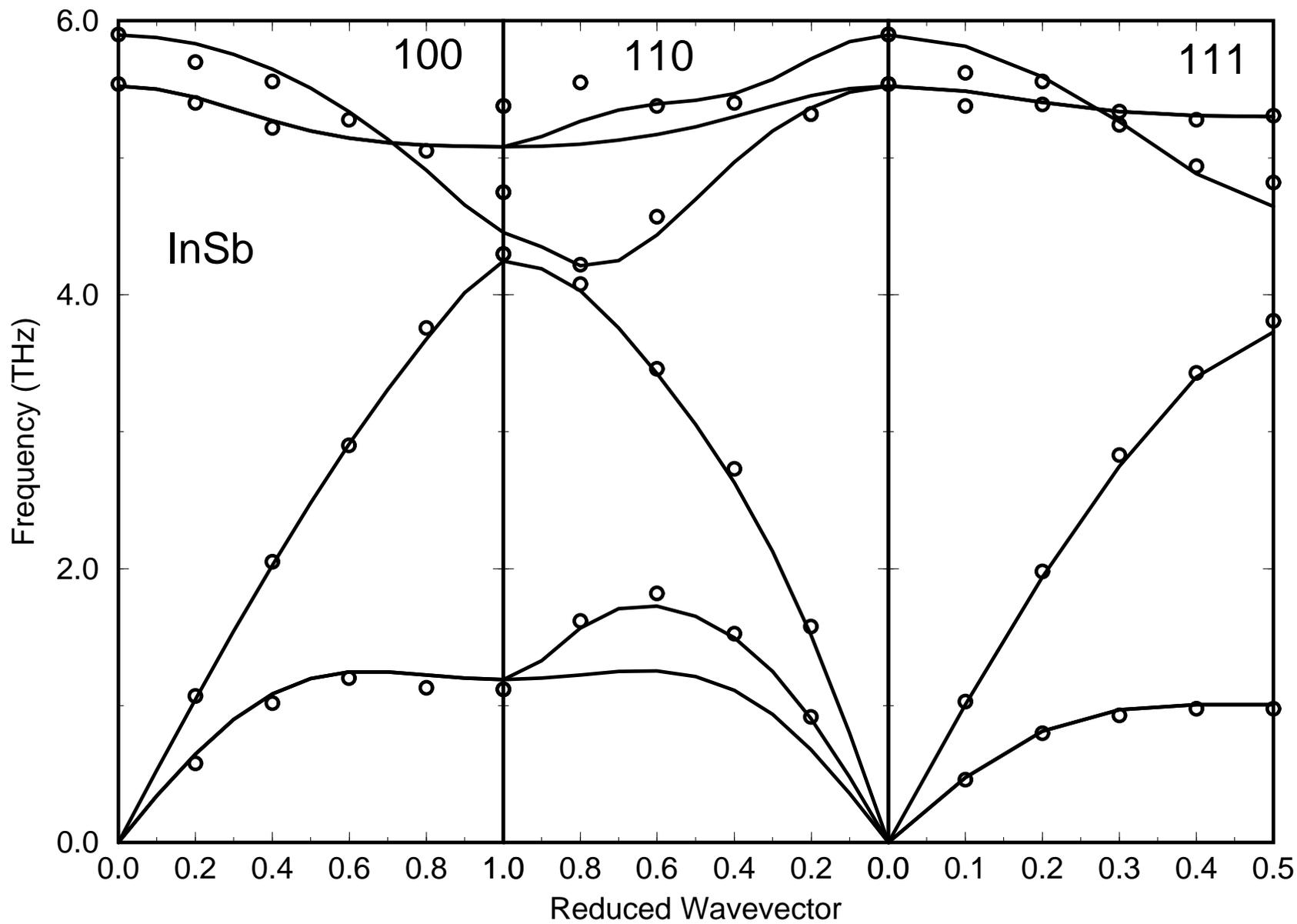
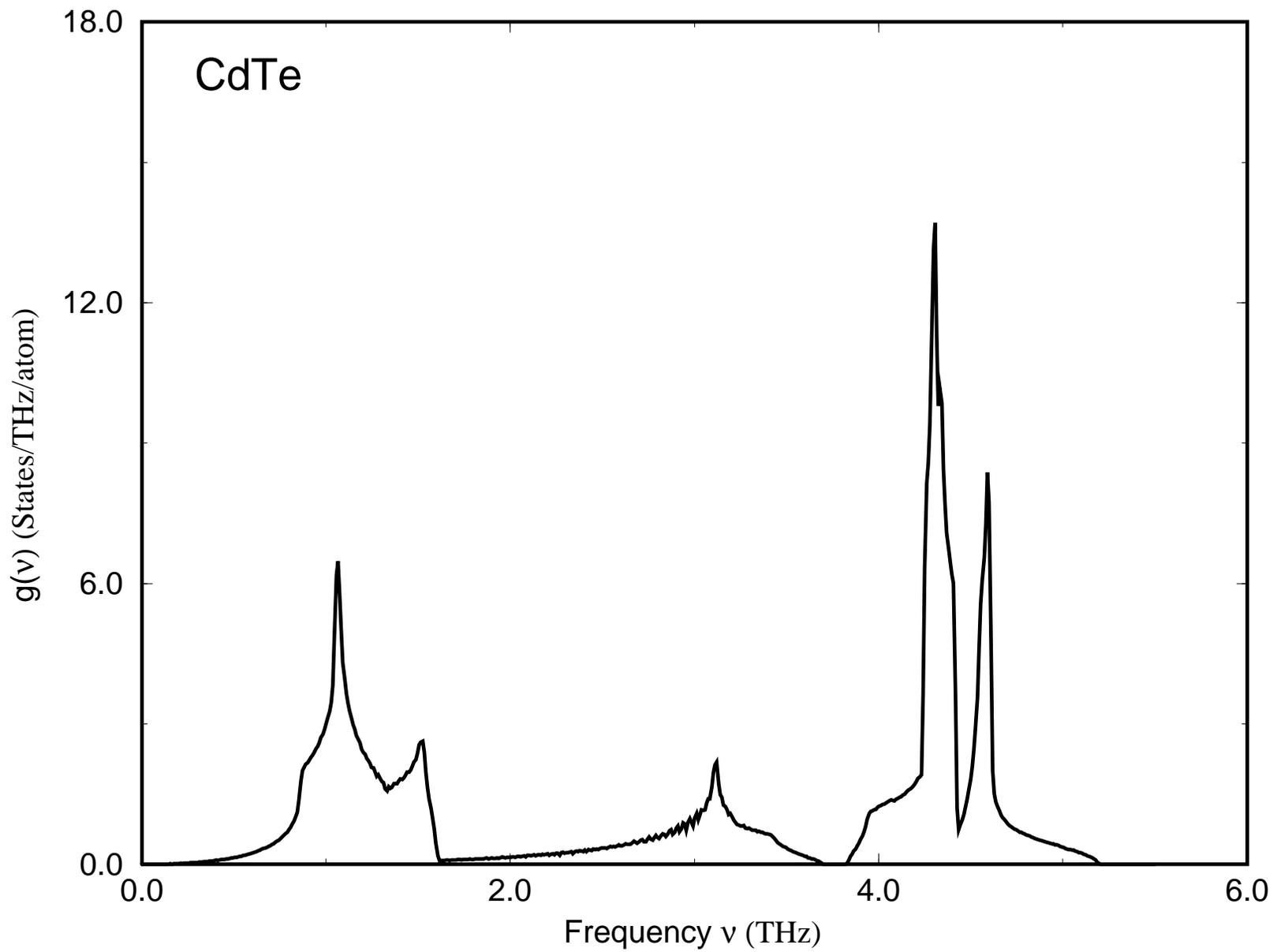
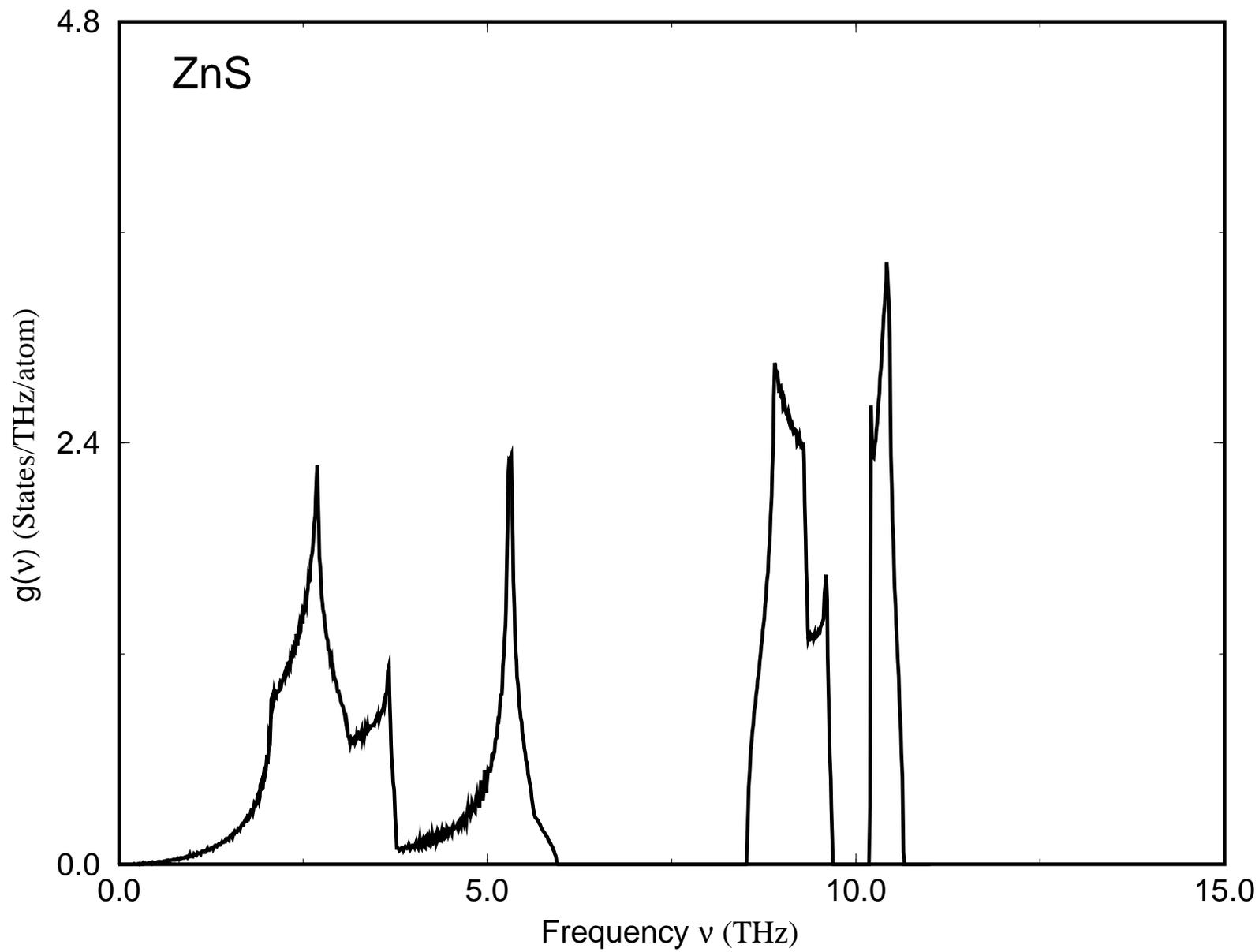
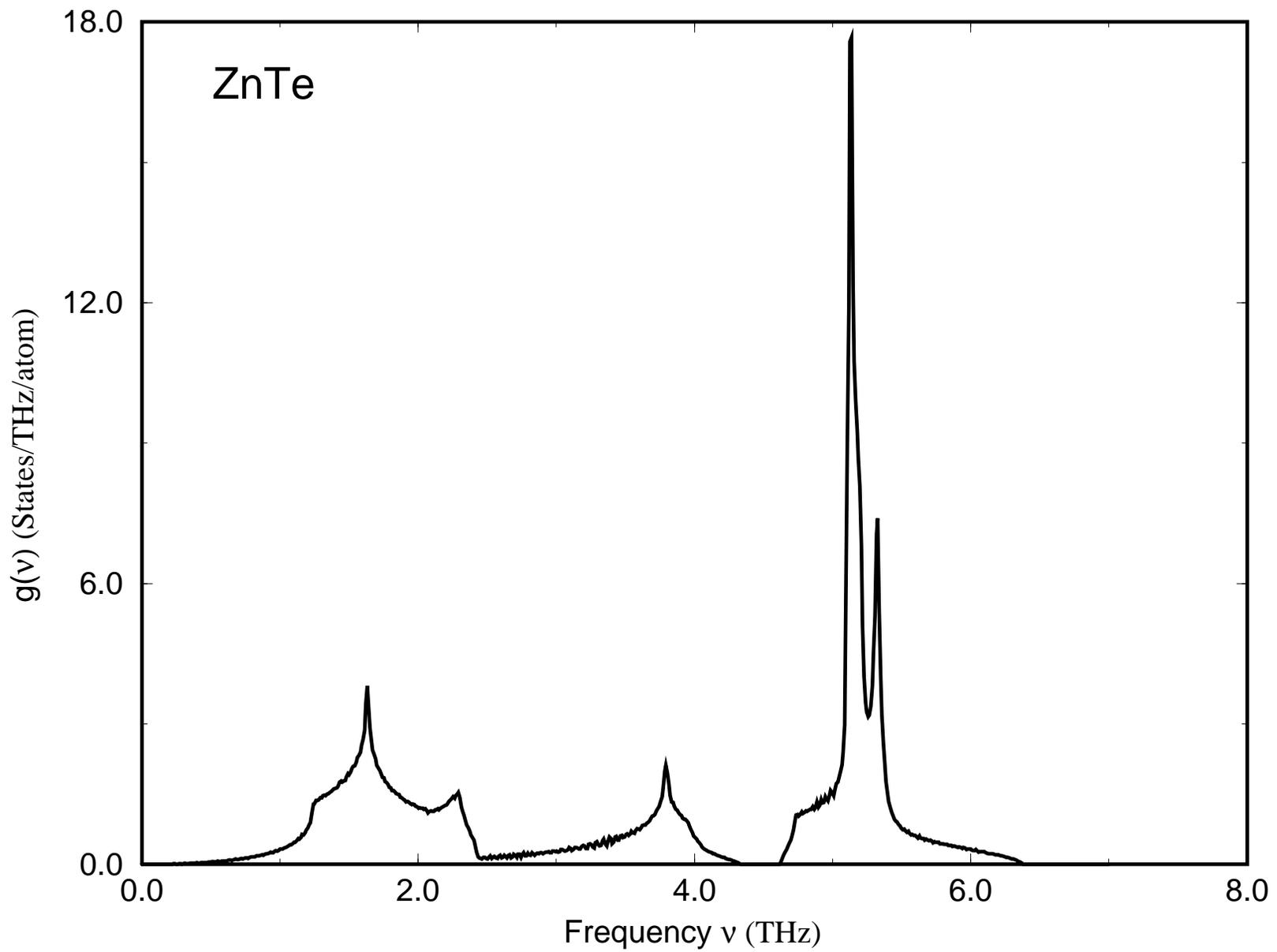
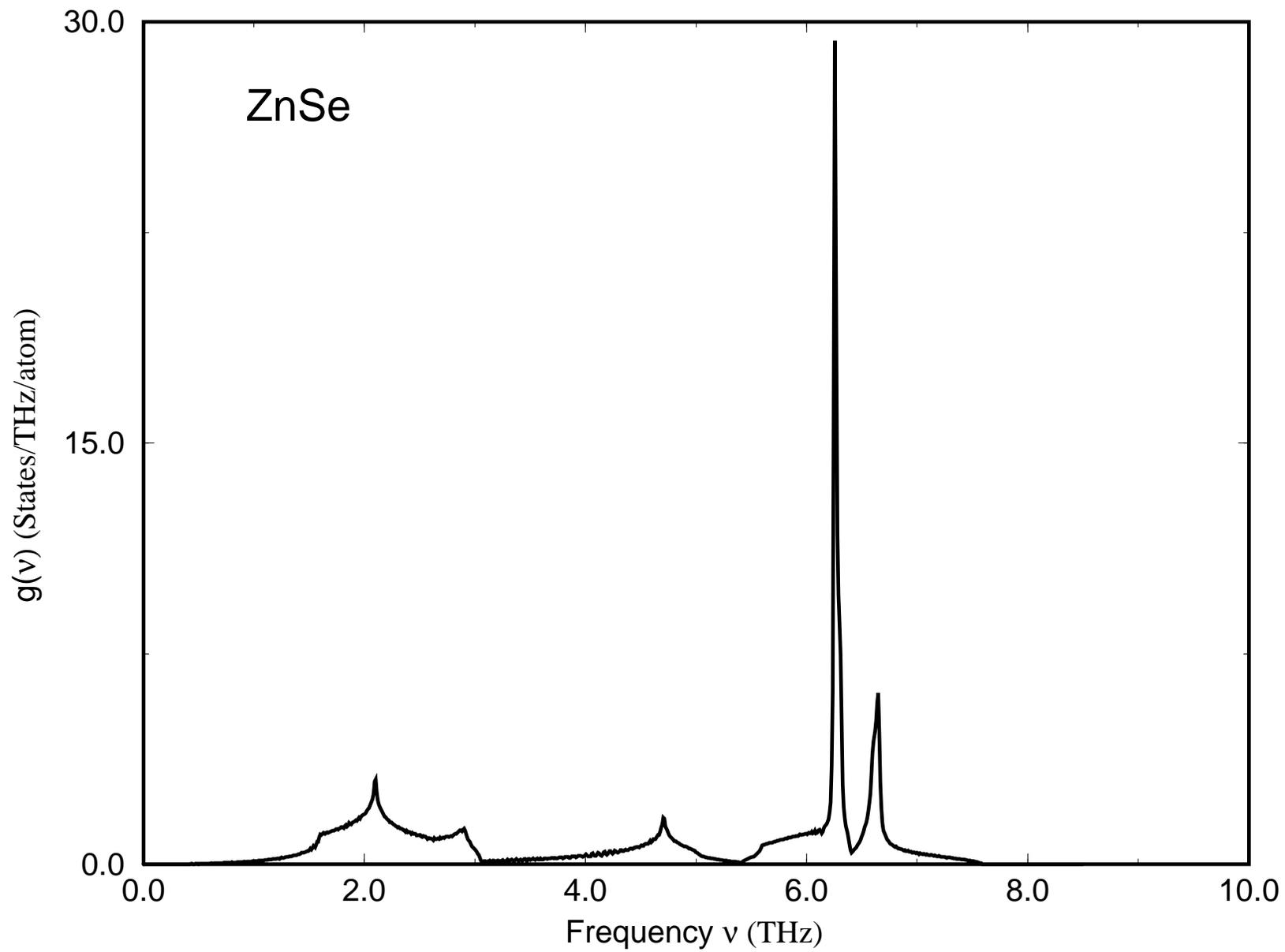
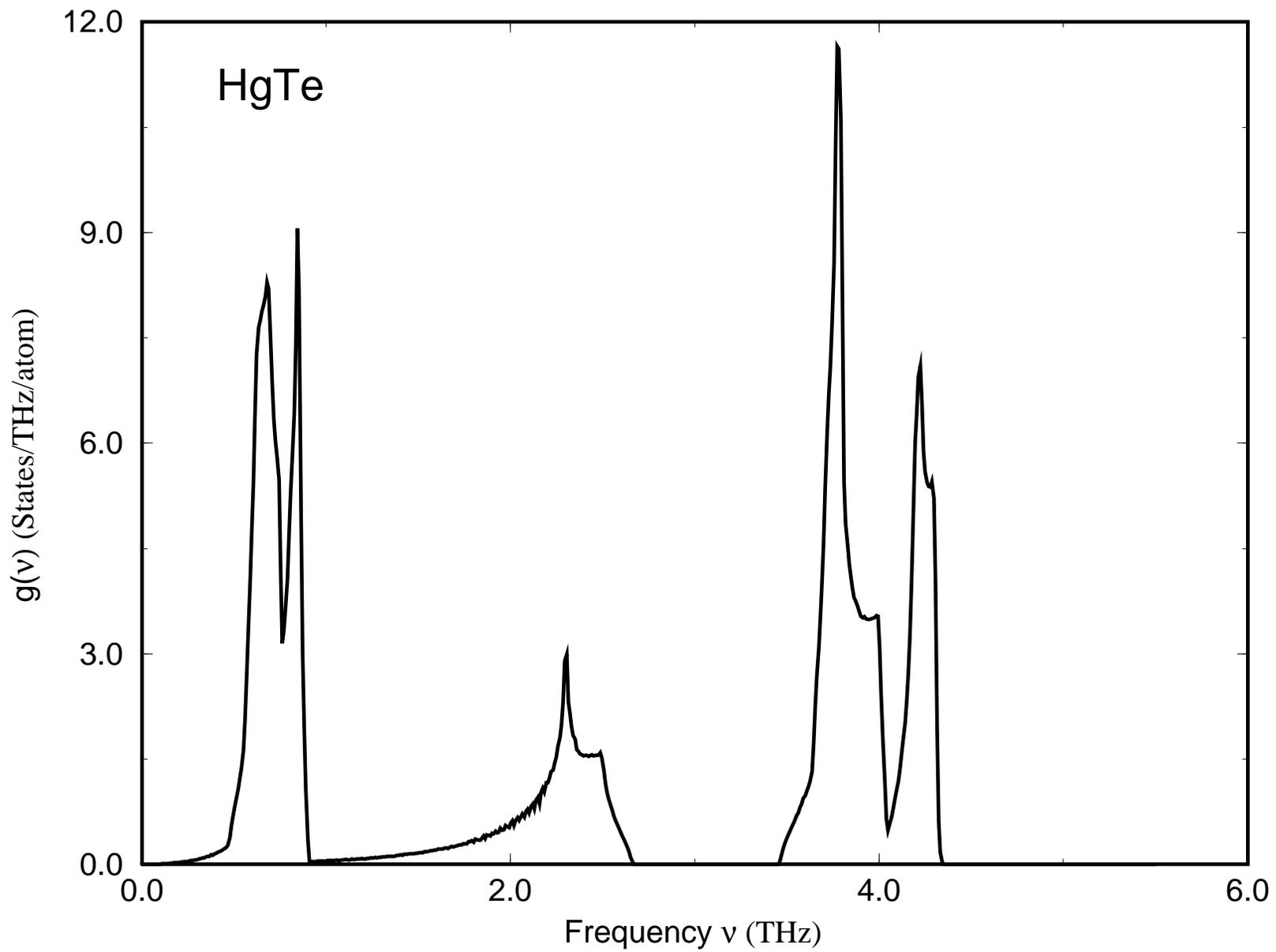
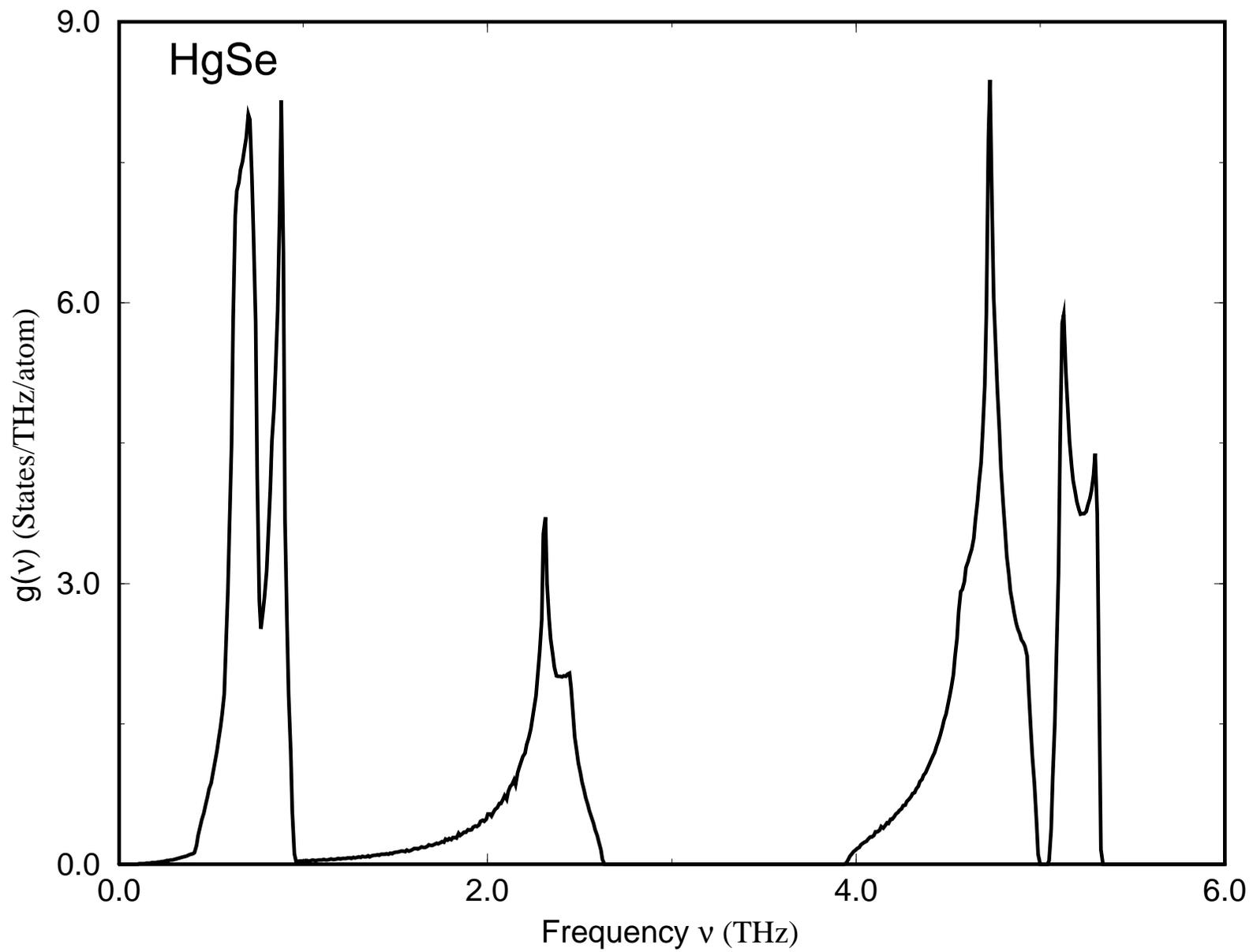
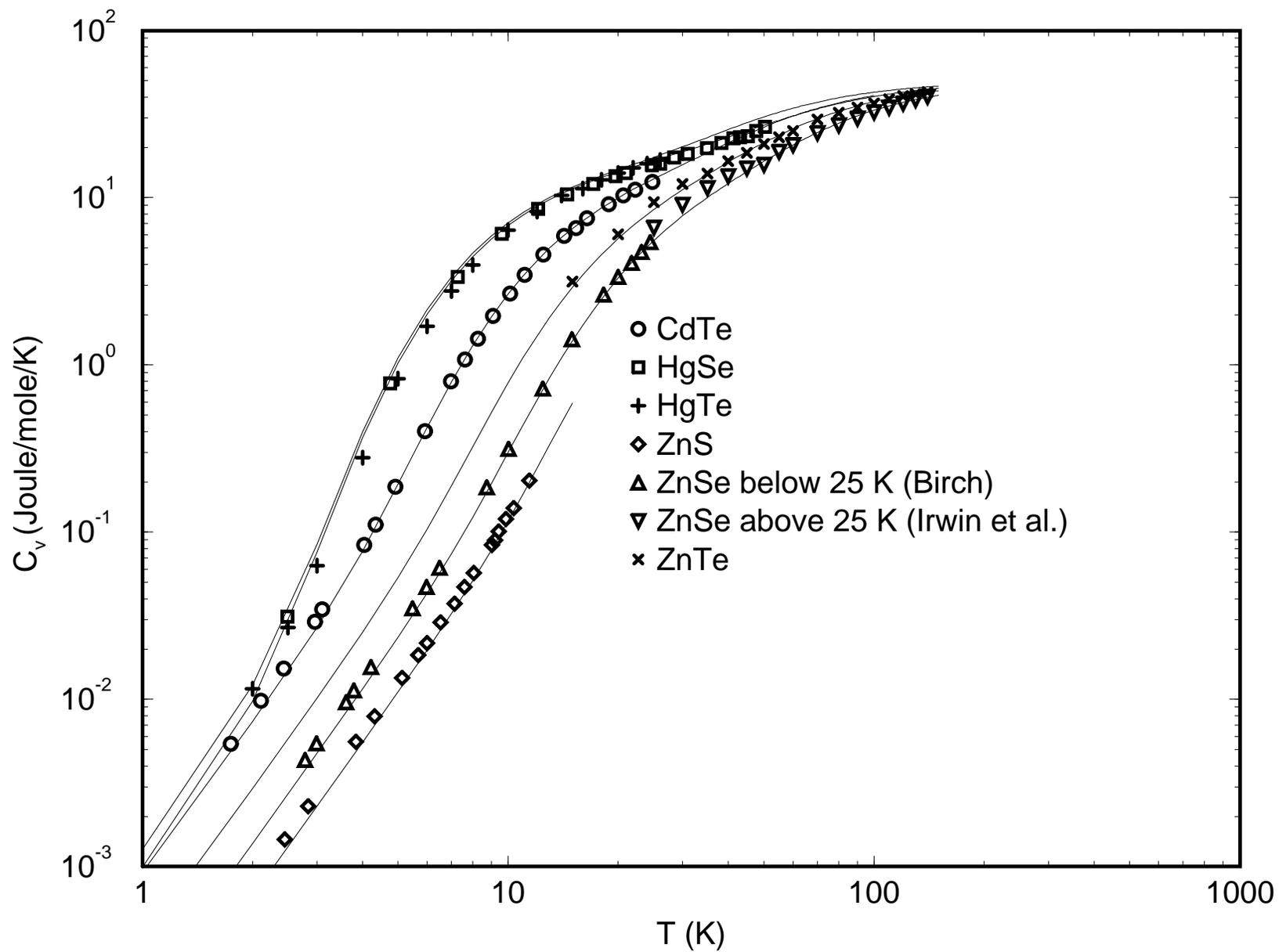
\end{document}